\documentclass{ws-ijmpa} 
\begin{document} 

\markboth{A. Gal}{$K^-d$ Scattering Length} 

%
\catchline{}{}{}{}{}
%

\title{ON THE SCATTERING LENGTH OF THE $K^-d$ SYSTEM\footnote{invited talk 
delivered at MESON 2006, 9th International Workshop on Meson Production, 
Properties and Interaction, Krakow, Poland, 9-13 June 2006.}} 

\author{AVRAHAM GAL} 

\address{Racah Institute of Physics, The Hebrew University \\ 
Jerusalem 91904, Israel \\ avragal@vms.huji.ac.il} 

\maketitle


\begin{abstract}
Multiple-scattering approximations to Faddeev calculations of the $K^-d$ 
scattering length are reviewed and compared with published 
$\bar K NN - \pi YN$ fully reactive Faddeev calculations. 
A new multiple-scattering approximation which goes beyond the `fixed-center' 
assumption for the nucleons is proposed, aiming at accuracies of $5-10\%$. 
A precise value of the $K^-d$ scattering length from the measurement of the 
$K^-d$ $1s$ atomic level shift and width, planned by the DEAR/SIDDHARTA 
collaboration, plus a precise value for the $K^-p$ scattering length by 
improving the $K^-p$ atom measurements, are essential for extracting 
the $K^-n$ scattering length, for resolving persistent puzzles in low-energy 
$\bar K N$ phenomenology and for extrapolating into $\bar K$-nuclear systems. 

\end{abstract}

\ccode{PACS numbers: 13.75.Jz; 21.45.+v; 25.80.Nv; 36.10.Gv}

\section{Introduction}
\label{sec:int}

\begin{figure}[pb]
\label{fig:raha} 
\centerline{\epsfig{file=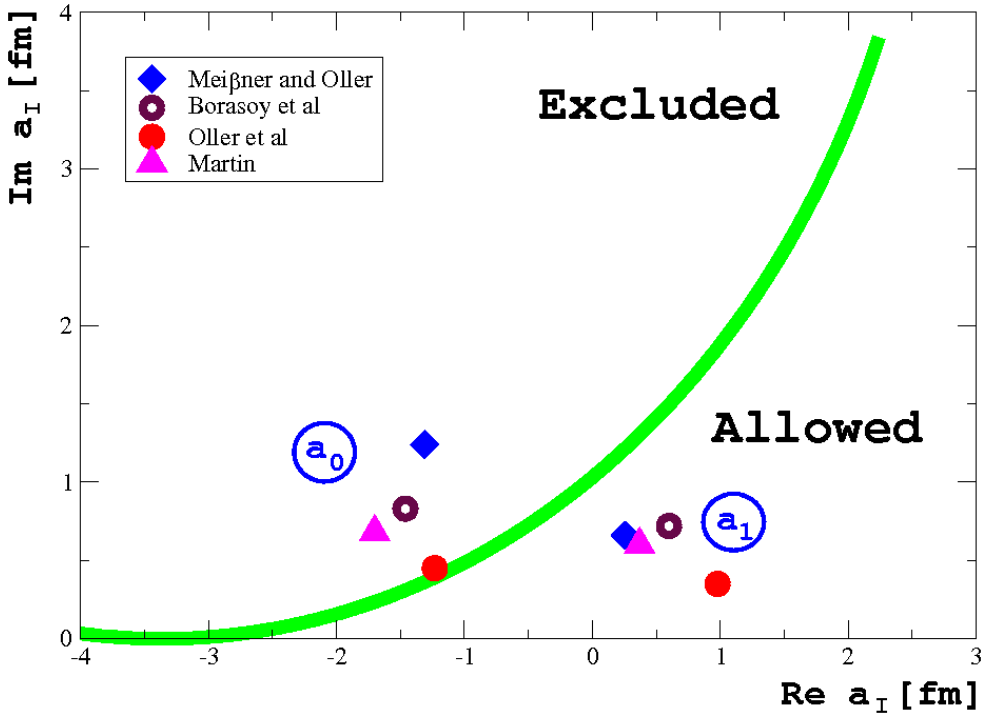,width=13cm}} 
\vspace*{8pt}
\caption{Excluded and allowed regions for $a_0,a_1$, 
from Mei{\ss}ner {\it et al.} Ref.~8,  
using the DEAR value for $a_{K^-p}$ in Eq.~(\ref{eq:split}). The values 
shown for $a_0$ from several calculations indicate a problem.} 
\end{figure}

The $\bar K N$ interaction near threshold is known, since the pioneering 
works by Dalitz and collaborators in the 1960s,\cite{Dal62} to be strongly 
attractive as well as strongly absorptive. 
The experimental data which traditionally have been used to constrain it 
consist of elastic, charge-exchange and 
$K^- p \rightarrow \pi Y$ reaction cross sections at low energies, 
as low as $p_{\rm lab}=100$~MeV/c, and of the $\Lambda(1405)$ resonance 
shape extracted from $\pi \Sigma$ final-state interaction below threshold. 
In addition, one has also three accurately determined branching ratios for 
$K^-$ absorption from rest. A crucial experimental datum near threshold, 
the (complex) $K^- p$ scattering length, is progressively becoming accurately 
determined in recent years from measurements of the energy shift and width of 
the $K^-p$ atomic $1s$ state.\cite{Iwa97,Bee05} 
The scattering length $a_{K^-p}$ may be expressed in terms of the $\bar K N$ 
$I=0,1$ scattering lengths $a_0$ and $a_1$, respectively, 
as follows\cite{DTu60}: 
\begin{equation} 
\label{eq:split} 
a_{K^-p} = \frac{\frac{1}{2}(a_0+a_1)+k_0a_0a_1}{1+k_0\frac{1}{2}(a_0+a_1)}~, 
\end{equation} 
where $k_0$ is the ${\bar K}^0 n$ cm momentum with respect to the $K^-p$ 
threshold. Using the NLO corrected Deser formula proposed recently by 
Mei{\ss}ner {\it et al.}\cite{MRR04} 
\begin{equation} 
\label{eq:Deser} 
\epsilon_{1s} - {\rm i} \frac{\Gamma_{1s}}{2} = -2\alpha^3\mu_{K^-p}^2a_{K^-p} 
(1-2\alpha\mu_{K^-p}({\rm ln}~\alpha - 1)a_{K^-p})~,
\end{equation} 
the value of $a_{K^-p}$ derived from the DEAR measurement\cite{Bee05}
\begin{equation} 
\label{eq:DEAR} 
a_{K^-p} = (-0.45 \pm 0.090) + {\rm i}(0.27 \pm 0.12)~{\rm fm} 
\end{equation} 
appears inconsistent with comprehensive fits to 
low-energy $K^- p$ scattering and reaction data, as discussed recently in 
Refs.~\refcite{BNW05,BMN06} and as shown pictorially in 
Fig.~\ref{fig:raha} taken from Ref.~\refcite{MRR06}. 
Furthermore, such fits leave the $K^- n$ scattering length $a_{K^-n}$
poorly determined. 

A precise measurement of the energy 
shift and width of the $K^-d$ atomic $1s$ state is called for, in order to 
determine $a_{K^-d}$, and this indeed is one of the prime aims of the 
DEAR/SIDDHARTA experimental collaboration (see e.g. Johann Zmeskal's talk 
in these Proceedings). A theoretically meaningful extraction of $a_{K^-n}$ 
from $a_{K^-p}$ and $a_{K^-d}$ would require a relatively handy and accurate 
multiple-scattering (MS) approximation to the more involved coupled-channel 
Faddeev calculation of the $K^-d$ system at low energy. Several Faddeev 
calculations of this sort have been reported.\cite{TGE81}\cdash\cite{BFM03} 
However, although the subject of MS approximations to the 
$K^-d$ scattering length has been discussed extensively in recent 
years,\cite{BDe99}\cdash\cite{WLo05} only {\it one} of these 
works produced a correct expression\cite{KOR01} within the commonly used 
`fixed-center' assumption for the nucleons, with a follow up recently in 
Ref.~\refcite{MRR06}.  

The main purpose of this work is to review the existing MS approximations 
and to propose a new MS approximation for $a_{K^-d}$. The proposed version 
goes beyond the `fixed-center' assumption. This should enable one 
to make a reasonable prediction for the $K^-d$ atom $1s$ level- shift and 
width on the basis of given values of $a_{K^-p}$ and $a_{K^-n}$ without 
having to perform new time-consuming Faddeev calculations or, 
{\it vice-versa}, to derive fairly reliably $a_{K^-n}$ from the joint 
measurement of the $1s$ level- shift and width in $K^-p$ and $K^-d$ atoms.

\section{Multiple-Scattering Approximations} 
\label{sec:MSA} 

It was shown in the first fully reactive $\bar K NN - \pi YN$ 
($Y=\Lambda,\Sigma$) Faddeev calculation\cite{TGE81} that $a_{K^-d}$ may 
be approximated to a few percent by neglecting the hyperonic channels 
altogether, in accordance with the conjecture by Schick and 
Gibson,\cite{SGi78} provided the input to the $\bar K NN$ 
Faddeev equations consists of $\bar K N$ t matrices which implicitly 
account for the $\bar K N - \pi Y$ coupling. 
The Faddeev equations for the $K^-pn$ 3-body system, with initial 
$K^- d$ configuration, assume the form 
\begin{equation} 
\label{eq:T} 
T_1 = t_1 + t_1 G_0 (T_2 + T_3), ~~T_2 = t_2 + t_2 G_0 (T_3 + T_1), 
~~T_3 = t_3 G_0 (T_1 + T_2)~, 
\end{equation} 
where $G_0$ is the free Green's function and the full $T$ matrix is given 
by its breakdown into partial $T$ matrices:~~$T=T_1+T_2+T_3$. The partial 
$T_1$ and $T_2$, upon a suitable choice of channels, stand for all the MS
terms that start with a $K^-p$ collision (2-body $t_1$ matrix) and with 
a $K^-n$ collision (2-body $t_2$ matrix), respectively, whereas $T_3$ 
consists of the terms that start with $pn$ collisions (2-body $t_3$ 
matrix). The choice of $K^-d$ initial state dictates that the last of 
Eqs. (\ref{eq:T}) does not have an inhomogeneous term $t_3$. 
No antisymmetrization is explicitly applied yet for the $NN$ subsystem. 
The $T$ matrix which reduces in the limit of zero kinetic energy to the 
$K^-d$ scattering length, up to a kinematical factor, is $T_{K^-d}=T_1+T_2$. 
We now outline several MS approximations, progressively by order of 
complexity, beginning with the Brueckner formula\cite{Bru53} which 
was motivated by $\pi d$ scattering for static nucleons. Charge-exchange 
degrees of freedom are then introduced and, wherever charge independence 
holds, the Brueckner formula is generalized in terms of isoscalar and 
isovector meson-nuclear scattering lengths.  

\subsection{Brueckner formula} 
\label{subsec:bru1} 

The Brueckner formula is obtained by neglecting $t_3$ and hence also $T_3$ 
in Eqs. (\ref{eq:T}), 
\begin{equation} 
\label{eq:Tfc} 
T_1 = t_1 + t_1 G_0 T_2~, ~~~T_2 = t_2 + t_2 G_0 T_1~, 
\end{equation} 
and treating the nucleons as fixed centers. 
Furthermore, mostly for the sake of illustration, the charge-exchange two-body 
channel $K^-p \to {\bar K}^0 n$ is switched off, leading 
to the following expression: 
\begin{equation} 
\label{eq:Tbru} 
T_{K^-d} = T_1 + T_2 = \frac{1}{1 - t_1 G_0 t_2 G_0}(t_1 + t_1 G_0 t_2) 
+ \frac{1}{1 - t_2 G_0 t_1 G_0}(t_2 + t_2 G_0 t_1)~.
\end{equation} 

For zero-range interactions, and considering the necessary kinematical 
factors to transform from $t$ matrices to scattering lengths $a$, it is 
straightforward to obtain from Eq.~(\ref{eq:Tbru}) the Brueckner formula: 

\begin{equation} 
\label{eq:int} 
a_{K^-d}={\left (1+\frac{m_K}{m_d} \right )}^{-1} 
\int{a_{K^-d}(r){|\psi_d({\bf r})|}^2 {\rm d}{\bf r}}~, 
\end{equation} 
\begin{equation} 
\label{eq:bru} 
a_{K^-d}(r) = \frac{{\tilde a}_p + {\tilde a}_n + 
2 {\tilde a}_p {\tilde a}_n /r} 
{1-{\tilde a}_p {\tilde a}_n /r^2}~, 
\end{equation} 
where $\tilde a = (1+m_K/m_N) a$, with $a$ standing generically for 
$a_{K^-p}=a_p$ and for $a_{K^-n}=a_n$ in the $K^-N$ cm system. 
The numerator in the Brueckner formula consists of single- and 
double-scattering terms, whereas the denominator provides for the 
renormalization of these terms to include all the higher-order scattering 
terms. The relevant expansion parameter is $\tilde a <1/r>_d$, 
where $<1/r>_d \approx 0.5~{\rm fm}^{-1}$ for the deuteron 
wavefunction. Hence, this series faces divergence once the scattering length 
is of order $a \sim 1$~fm or more, as we encounter for the $I=0~\bar K N$ 
channel. In contrast, for low-energy pions\footnote{for pions change 
everywhere $m_K \to m_{\pi}$.} $a \sim 0.1$~fm and the single- and 
double-scattering terms (augmented by charge-exchange scattering as done 
below) provide an excellent approximation to the Brueckner formula for the 
$\pi^-d$ scattering length.\cite{Del01} The generalization of the 
Brueckner formula to include the charge exchange reaction 
on the proton (on-shell for $\pi^-$, off-shell for $K^-$) is given below. 

\subsection{Including charge exchange in the Brueckner formula} 
\label{subsec:bru2} 

The inclusion of charge-exchange degrees of freedom for $K^-d$ scattering 
at threshold is due to Ref.~\refcite{KOR01}: 
\begin{equation} 
\label{eq:brux} 
a_{K^-d}(r) = \frac{{\tilde a}_p + {\tilde a}_n +
(2 {\tilde a}_p {\tilde a}_n - b_x^2)/r - 2 b_x^2 {\tilde a}_n /r^2} 
{1 - {\tilde a}_p {\tilde a}_n /r^2 + b_x^2 {\tilde a}_n /r^3}~, 
\end{equation} 
where $b_x^2 = {\tilde a}_x^2/(1+{\tilde a}_u/r)$, with $a_x$ and $a_u$ 
the threshold scattering amplitudes for $K^-p \rightarrow {\bar K}^0n$ 
and ${\bar K}^0n \rightarrow {\bar K}^0n$ respectively. The derivation of 
Eq.~(\ref{eq:brux}) shows that it holds also for $a_{\pi^-d}$, provided 
that the replacement $(K^-,{\bar K}^0)\rightarrow(\pi^-,\pi^0)$ is made 
in the $\bar K N$ amplitudes. The charge-exchange scattering contribution 
to $a_{\pi^-d}$ as well as to $a_{K^-d}$ is substantial. 
When charge independence is assumed, all four $\bar K N$ amplitudes in 
Eq.~(\ref{eq:brux}) are given in terms of the isoscalar and isovector 
threshold amplitudes $b_0, b_1$ which are related to the isospin scattering 
lengths $a_0, a_1$ by\footnote{the isospin basis may still be used when 
charge independence is only broken by different $\bar K N$ 
thresholds.\cite{MRR06}} 
\begin{equation} 
\label{eq:ampK} 
b_0 + b_1 {\vec{\tau}}_{\bar K}\cdot{\vec{\tau}}_N~:~~~~~~ 
b_0=\frac{1}{4}(3a_1+a_0)~,~~~b_1=\frac{1}{4}(a_1-a_0)~. 
\end{equation} 
Eq.~(\ref{eq:brux}) simplifies then to 
\begin{equation} 
\label{eq:iso} 
a_{K^-d}(r) = \frac{2{\tilde b}_0 - 2({\tilde b}_0 + {\tilde b}_1) 
(3{\tilde b}_1 - {\tilde b}_0)\frac{1}{r}}{1 - 2{\tilde b}_1 \frac{1}{r} 
+ ({\tilde b}_0 + {\tilde b}_1)(3{\tilde b}_1 - {\tilde b}_0) \frac{1}{r^2}} = 
\frac{\frac{1}{2}(3{\tilde a}_1 + {\tilde a}_0) + 
2 {\tilde a}_0 {\tilde a}_1 \frac{1}{r}} 
{1 - \frac{1}{2}({\tilde a}_1 - {\tilde a}_0)\frac{1}{r} - 
{\tilde a}_0 {\tilde a}_1 \frac{1}{r^2}}~, 
\end{equation} 
in close analogy to Deloff's expression\cite{Del01} for $\pi^-d$ scattering 
at threshold: 
\begin{equation} 
\label{eq:pid} 
a_{\pi^-d}(r) = \frac{ 2{\tilde b}_0 - 2({\tilde b}_0 + {\tilde b}_1) 
(2{\tilde b}_1 - {\tilde b}_0)\frac{1}{r}}{1 - {\tilde b}_1 \frac{1}{r} 
+ ({\tilde b}_0 + {\tilde b}_1)(2{\tilde b}_1 - {\tilde b}_0)\frac{1}{r^2}} = 
\frac{\frac{2}{3}(2{\tilde a}_{\frac{3}{2}} + {\tilde a}_{\frac{1}{2}}) + 
2{\tilde a}_{\frac{1}{2}} {\tilde a}_{\frac{3}{2}} \frac{1}{r}} 
{1 - \frac{1}{3}({\tilde a}_{\frac{3}{2}} - 
{\tilde a}_{\frac{1}{2}})\frac{1}{r} - 
{\tilde a}_{\frac{1}{2}} {\tilde a}_{\frac{3}{2}} \frac{1}{r^2}}~, 
\end{equation} 
where the isoscalar and isovector $\pi N$ threshold amplitudes are now given 
in terms of the isospin scattering lengths $a_{1/2}, a_{3/2}$ by 
\begin{equation} 
\label{eq:ampi} 
b_0 + b_1 {\vec t}_{\pi}\cdot{\vec {\tau}}_N~:~~~~~~ 
b_0 = \frac{1}{3}(2a_{3/2} + a_{1/2})~,~~~
b_1 = \frac{1}{3}(a_{3/2} - a_{1/2})~. 
\end{equation} 
Eqs.~(\ref{eq:iso},\ref{eq:pid}) may be obtained directly from 
Eq.~(\ref{eq:Tbru}) treating it as an operator relationship in isospace. 

\begin{table}[ph]
\tbl{Comparison of values of $a_{K^-d}$ (in units of fm) from cited Faddeev 
calculations with values from the `fixed center' MS expression 
Eqs.~(\ref{eq:int},\ref{eq:brux}) for `particle' and 
Eqs.~(\ref{eq:int},\ref{eq:iso}) for `isospin', under one further 
approximation, see text. 
The value of $a_{K^-d}$ listed for the Faddeev calculation of 
Toker {\it et al.} Ref.~9 replaces the value given there at 
$p_{K^-}=15$~MeV/c.} 
{\begin{tabular}{@{}cccccc@{}} \toprule 
Ref. & basis & $a_{K^-p}$ & $a_{K^-n}$ & Fad $a_{K^-d}$ & MS $a_{K^-d}$ \\ 
\colrule 
TGE\cite{TGE81} & isospin  & $-0.655+{\rm i}~0.705$ & $0.350+{\rm i}~0.660$ & 
$-1.67+{\rm i}~1.00$ & $-1.46+{\rm i}~1.09$ \\ 
TDD\cite{TDD86} & isospin  & $-0.645+{\rm i}~0.725$ & $0.320+{\rm i}~0.700$ & 
$-1.34+{\rm i}~1.04$ & $-1.42+{\rm i}~1.09$ \\ 
BFMS\cite{BFM03}& particle & $-0.888+{\rm i}~0.867$ & $0.544+{\rm i}~0.644$ & 
$-1.80+{\rm i}~1.55$ & $-1.73+{\rm i}~1.06$ \\ 
\botrule 
\end{tabular} \label{tab:fad}} 
\end{table} 

In Table~\ref{tab:fad} we show the quality of the `fixed center' MS 
approximation. Eqs.~(\ref{eq:int},\ref{eq:iso}) were used to compare with 
values for $a_{K^-d}$ reported in Faddeev calculations done in the isospin 
basis, with the further approximation of replacing $r^{-n}$ by the TPE matrix 
elements $<r^{-n}>$ for $n=1,2$ from Table 1 in Ref.~\refcite{VAr06}. 
The values of $a_p$ and $a_n$ listed in the table are sufficient to determine 
the input scattering lengths $a_0$ and $a_1$ for Eq.~(\ref{eq:iso}). This  
version of MS approximation reproduces the real part of $a_{K^-d}$ to better 
than about $15\%$ and the imaginary part to within $5-10\%$. 
For the particle-basis Faddeev calculation of Ref.~\refcite{BFM03}, 
Eqs.~(\ref{eq:int},\ref{eq:brux}) were used. The extra scattering lengths, 
beyond $a_p$ and $a_n$, are not listed here. The renormalized TPE value for 
$<r^{-3}>$ was used.\cite{VAr06} The resulting value of $a_{K^-d}$ is 
sensitive to the value used for this matrix element which diverges unless 
renormalized; scaling it down from $<r^{-2}>$ by the ratio of 
$<r^{-2}>/<r^{-1}>$, we get $a_{K^-d}=-1.96+{\rm i}1.37$~fm, in better 
agreement with the Faddeev value~~$-1.80+{\rm i}1.55$~fm.\cite{BFM03}  

In Table~\ref{tab:MS} we list two MS evaluations of $a_{K^-d}$ 
using a deuteron wavefunction based on the Paris potential and $\bar K N$ 
input which is similar to that used in the TDD\cite{TDD86} Faddeev 
calculation. The accuracy of these MS evaluations is typically $25\%$. 
Also listed is our own MS approximation (from Table~\ref{tab:fad}) which, 
perhaps fortuitously, works to about $5\%$. 
 
\begin{table}[ph]
\tbl{MS approximations for a Faddeev calculation of $a_{K^-d}$ 
(values in fm).}  
{\begin{tabular}{@{}cccc@{}} \toprule 
Faddeev\cite{TDD86} & MS\cite{KOR01} & MS\cite{MRR06} & present MS \\
\colrule 
$-1.34+{\rm i}~1.04$ & $-1.54+{\rm i}~1.29$ & $-1.66+{\rm i}~1.28$ & 
$-1.42+{\rm i}~1.09$ \\ 
\botrule 
\end{tabular} \label{tab:MS}} 
\end{table}

\subsection{Beyond `fixed centers'} 
\label{subsec:bru3}
 
Under the `fixed center' assumption we have suppressed the terms $t_1 G_0 T_3$ 
and $t_2 G_0 T_3$ in the coupled Faddeev equations for $T_1$ and $T_2$, 
respectively, in Eq.~(\ref{eq:T}). These terms may be rewritten in the form 
\begin{equation} 
\label{eq:T13} 
t_1 G_0 T_3 = t_1 G_0 t_3 G_0 (T_1 + T_2) = t_1 (\Delta G_0) (T_1 + T_2)~, 
\end{equation} 
\begin{equation} 
\label{eq:T23} 
t_2 G_0 T_3 = t_2 G_0 t_3 G_0 (T_1 + T_2) = t_2 (\Delta G_0) (T_1 + T_2)~,  
\end{equation} 
where $\Delta G_0 = G_3 - G_0 = G_0 t_3 G_0$, so that $G_3$ is a Green's 
function which takes full account of the $NN$ interaction but is still free 
with respect to the meson-nucleon interactions. This leads to the following 
coupled equations for $T_1$ and $T_2$: 
\begin{equation} 
\label{eq:T'} 
\left( 1- t_1 (\Delta G_0) \right) T_1 = t_1 + t_1 G_3 T_2~,~~~
\left( 1- t_2 (\Delta G_0) \right) T_2 = t_2 + t_2 G_3 T_1 ~, 
\end{equation} 
which are in the form of the `fixed center' Faddeev equations (\ref{eq:Tfc}) 
with $G_0$ replaced by $G_3$ and with {\it renormalized} $t_1$ and $t_2$: 
\begin{equation} 
\label{eq:t'} 
t'_j = \left( 1- t_j (\Delta G_0) \right)^{-1} t_j 
= t_j + t_j (\Delta G_0) t_j + \cdots ~,~~~~~~ j=1,2~. 
\end{equation} 
We thus obtain the following improvement over the Brueckner formula 
Eq.~(\ref{eq:Tbru}): 
\begin{equation} 
\label{eq:T'bru}
T_{K^-d} = T_1 + T_2 = \frac{1}{1 - t'_1 G_3 t'_2 G_3}(t'_1 + t'_1 G_3 t'_2) 
+ \frac{1}{1 - t'_2 G_3 t'_1 G_3}(t'_2 + t'_2 G_3 t'_1)~.
\end{equation} 
This expression for $T_{K^-d}$ provides as transparent and applicable MS 
expansion, with $t_j \rightarrow t'_j~,~~j=1,2~,$ and $G_0 \rightarrow G_3$, 
as the Brueckner formula MS expansion is. Charge exchange degrees of freedom 
can be introduced in a straightforward manner, as done in 
subsection~\ref{subsec:bru2}. It is conceivable that one can reach in this way 
a level of accuracy of $5-10\%$ for approximating fully-reactive Faddeev 
calculations.

\section{Discussion and conclusions} 
\label{sec:disc} 

In this talk I have surveyed schematically the derivation of 
MS approximations to the few available fully-reactive Faddeev calculations for 
$a_{K^-d}$ under the generally accepted practice of treating the nucleons 
as fixed centers. These `fixed center' MS approximations, which provide 
an excellent approximation scheme already at the double-scattering order 
for $a_{\pi^-d}$, require the full summation of the MS series for $a_{K^-d}$, 
using Eq.~(\ref{eq:brux}) in the particle basis or Eq.~(\ref{eq:iso}) in the 
isospin basis. The accuracy thus provided is in the range of $10-25\%$. 
It should be stressed that Effective Field Theory approaches do not yet offer 
any alternative scheme at present.\cite{MRR06} 
I have subsequently outlined a theoretical MS approach for going beyond the 
`fixed center' assumption while keeping the formal appearance of the 
`fixed center' MS formulae, at the expense of renormalizing the input 
$\bar K N$ scattering amplitude and the free Green's function which for 
fixed centers gives rise to the $r^{-n}$ dependence of the MS series terms. 
It is conceivable that one can reach in this way a level of accuracy of 
$5-10\%$ for approximating fully-reactive Faddeev calculations. 
The aim of this project is to free oneself of depending on available 
Faddeev calculations, which might become obsolete if the DEAR 
measurement\cite{Bee05} of the $K^-p$ atomic $1s$ level shift and width is 
confirmed and the error bars further reduced, since {\it all} the published 
Faddeev calculations of $a_{K^-d}$ use considerably larger values for the 
$a_{K^-p}$ input. 

\section*{Acknowledgments} 
This work was supported in part by grant 757/05 of the Israel Science 
Foundation, Jerusalem, Israel.

\end{document}